\documentclass[aps,prl,reprint]{revtex4-1}

\usepackage{xfrac} % bold math symbols
\usepackage{blindtext}
\usepackage{graphicx,color,amsmath,amssymb}
\usepackage{dcolumn}% Align table columns on decimal point
\usepackage{bm}% bold math

\begin{document}

\title{Self-consistent description of electrokinetic phenomena in particle-based simulations}

\author{Juan P. Hern\'{a}ndez-Ortiz}
%\email{jphernandezo@unal.edu.co}
%\affiliation{Departamento de Materiales, Universidad Nacional de Colombia, Carrera 80 \# 65-223, Bloque M3-050, Medell\'in Colombia}
\affiliation{Institute for Molecular Engineering, University of Chicago, Chicago, Illinois 60637, United States}

\author{Juan J. de Pablo}
\email{depablo@uchicago.edu}
\affiliation{Institute for Molecular Engineering, University of Chicago, Chicago, Illinois 60637, United States}
	
\date{\today}

\begin{abstract}
A new computational method is presented for study suspensions of charged soft particles undergoing fluctuating hydrodynamic and electrostatic interactions. 
The proposed model is appropriate for polymers, proteins and porous particles embedded in a
continuum electrolyte. A self-consistent Langevin description of the particles is adopted in which hydrodynamic and electrostatic interactions are
included through a Green's function formalism. An Ewald-like split is adopted in order to satisfy arbitrary boundary conditions for the Stokeslet and Poisson Green functions, thereby providing a formalism that is applicable to any geometry and that can be extended to deformable objects.
The convection-diffusion equation for the continuum ions is solved simultaneously considering Nernst-Planck diffusion.
The method can be applied to systems at equilibrium and far from equilibrium. Its applicability is demonstrated in the context of electrokinetic motion, where it is shown that the ionic clouds associated with individual particles can be severely altered by the flow and concentration, leading to intriguing cooperative effects. 
\end{abstract}

\pacs{}

\keywords{Electrokinetic phenomena, soft-particles, hydrodynamic and electrostatic interactions}

\maketitle

% INTRO
There is considerable interest in understanding the structure and dynamics of suspensions of charged particles over multiple 
length scales, both at equilibrium and far from equilibrium. 
Examples include DNA or protein flow in mocrofluidic devices, in cellular environments, or colloidal self-assembly in external fields. 
Beyond any direct interaction (van der Waals or electrostatic) between particles, the motion of a particle in solution induces 
important hydrodynamic and electrostatic interactions that some times compete against each other, leading to 
electro-osmotic or electro-kinetic phenomena that remain poorly understood. 

\begin{figure}
\begin{center}
\includegraphics[width=0.48\textwidth]{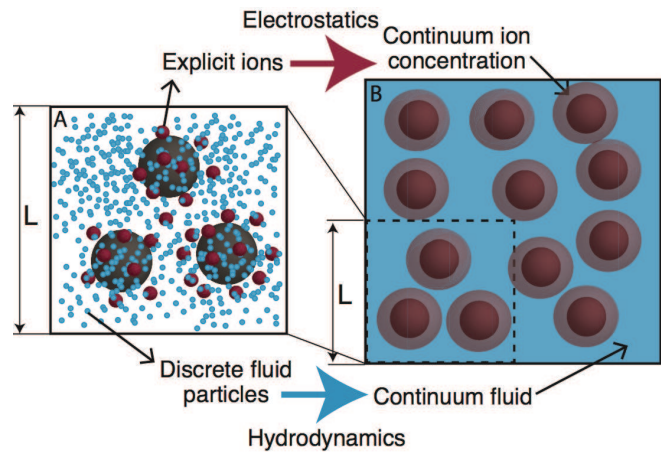}
\caption{Discrete charged soft-particles embedded in an electrolyte solvent. 
(A) A level of description that requires to treat soft-particles, ions and fluid as discrete entities. (B) The proposed method 
provides resolution for the interesting discrete entities whereas a continuum description for the electrolyte solvent.  }
\label{fig:cartoon}
\end{center}
\end{figure}

Resolving the dynamics of solvents or charged species over short and long length scales remains a significant 
challenge. The central question is how to evolve these systems, while preserving molecular resolution of 
discrete macromolecules or colloids, adopting continuum descriptions for solvent and ions (see Figure~\ref{fig:cartoon}). 
Past attempts have primarily relied on 
Lattice Boltzmann (LB)~\cite{Melchionna:2004p969}, 
Stochastic Rotational Dynamics (SRD)~\cite{Padding:2006p475,Webster:2005p467},
Stokesian Dynamcis (SD)~\cite{BanchioBrady03}, 
Ewald sums~\cite{hasimoto_1959a} and 
the General geometry Ewald-like method (GgEm)~\cite{hernandez-ggem} 
for the hydrodynamic evolution, i.e. the momentum equations. 
On the other hand, solution of charges have been treated 
by Ewald-based methods~\cite{BoardEwaldReview96} or
Poisson-Boltzmann solutions~\cite{Roux:2008p3515,Im:2002p3512}. 
A notable exception, where both interactions are solved simultaneously, is the
Smoothed Profile Method (SPM)~\cite{Kim:2006io,Luo2010}. In SPM, 
discrete particles are included into the continuum formalism through smoothing functions 
(see the review of particle-based methods by T. Yamaguchi et. al~\cite{Yamaguchi:2010wr}). 
LB and SRD methods rely on collision operators to evolve fluid dynamics, thus imposing theoretical limits at zero 
inertia or strict incompressibility ($Ma = Re = 0$). 
In addition, they present limitations regarding the size of the systems that can be studied. 
A salient limitation of SPM and LB is the mesh dependency that naturally develops at finite concentrations of the discrete entities. 
The method that we propose in this work, provides a way of resolving these problems. 

We are presenting a novel theoretical method that resolves the dynamical coupling between discrete charged soft-particles and 
continuum electrolyte solvents. It is presented as a generalization of the GgEm approach, which can be used for bulk and confined
systems, both at equilibrium and far from equilibrium. 
The dynamics of the soft-particles follows a Fokker-Planck equation for the probability density, resulting in a Stochastic evolution 
equation. The solvent and the continuum ions, on the other hand, are evolved according to momentum and mass and balances, including 
Nernst-Planck diffusion mechanisms (see Fig.~\ref{fig:cartoon}). 

We consider a collection of $N_P$ soft-particles, with charge $ez_\nu$ and hydrodynamic radius $a$, suspended in a
solvent that includes $N_I$ charged species (continuum ions). The particles are ``soft"
in the sense that they are permeable to the continuum ions and the fluid flow. This model
is suitable for coarse-grained models of polymers, proteins and other biological systems.
Soft-particles and continuum ions contribute to a charge density, defined as
\begin{equation}
\rho({\bf x})  = F\sum_{j=1}^{N_I} z_jC_j({\bf x}) + \sum_{\nu=1}^{N_P} z_\nu e \delta({\bf x}- {\bf x}_\nu),
\end{equation}
where $F=eN_A$ is Faraday's constant ($N_A$ is Avogadro's number), $z_j$ is the valence of the continuum ions ($j=1,...,N_I$),
$C_j$ is the concentration of the continuum species,
$z_\nu$ is the soft-particle's valence ($\nu = 1,...,N_P$) and $e$ is the elementary charge. Where 
the soft-particles, at this point, are considered point-charges.
If electroneutrality is not satisfied at a local level,
the charge density will drive an electrostatic potential given
by the solution of Poisson's equation:
\begin{equation}
\nabla^2\phi({\bf x}) = - \rho({\bf x})/\epsilon_0\epsilon,
\label{eq:poisson}
\end{equation}
where $\epsilon_0$ is the vacuum permittivity and $\epsilon$ is the solvent relative permittivity.
The electric field ($\mathbf{E}({\bf x})=-\nabla\phi({\bf x})$) drives an electric force on the ions.
Electric forces on the continuum ions and the total forces on the soft-particles define a force density given by
\begin{equation}
\boldsymbol{\rho}({\bf x}) = F\sum_{j=1}^{N_I} z_jC_j({\bf x}){\bf E}({\bf x}) +
\sum_{\nu=1}^{N_p} {\bf f}_\nu\delta({\bf x}-{\bf x}_\nu),
\end{equation}
where $\mathbf{f}_\nu$ represents the total non-Brownian and non-hydrodynamic force acting on particle $\nu$.
Neglecting inertia ($Re=0$), the solvent velocity can be written as ${\bf v}({\bf x}) = {\bf v}_0({\bf x}) + {\bf u}({\bf x})$, (where ${\bf v}_0$({\bf x}) is the unperturbed velocity and ${\bf u}({\bf x})$ is the
velocity perturbation), and it is given by the solution of a Stokes system of equations. The velocity
perturbation is driven by the force density, i.e.
\begin{equation}
-\nabla p({\bf x}) + \eta \nabla^2 {\bf u}({\bf x}) = - \boldsymbol{\rho}({\bf x}), \qquad
 \nabla\cdot{\bf u}({\bf x}) = 0,
\label{eq:stokes}
\end{equation}
where $\eta$ is the solvent viscosity.

The evolution of the ions within the solvent follows a species balance with
the total flux defined as a sum of convection and diffusion fluxes. The latter are given by the
Nernst-Planck diffusion, resulting in:
\begin{equation}
\begin{split}
\frac{\partial C_j}{\partial t} = & - {\bf v}\cdot \nabla C_j + D_j\nabla^2 C_j + \\
 & D_j z_j (e/k_BT)\left[ C_j \nabla^2 \phi + \nabla C_j \cdot \nabla \phi \right],
\end{split}
\label{eq:mass-2}
\end{equation}
where $D_j$ is the diffusion coefficient of ion $j$, $k_B$ is Boltzmann's constant,
and $T$ is the absolute temperature.

Each of the $N_P$ discrete ions behaves, for the moment, as a point-force and a point-charge.
The discrete ions have a hydrodynamic radius $a$, and interact via a repulsive excluded volume potential.
Neglecting inertia (Reynolds, $Re=0$), for each discrete ion, $\nu=1,...,N_P$, the force balance requires
$\mathbf{f}_\nu^H+\mathbf{f}_\nu^S+\mathbf{f}_\nu^{E}+\mathbf{f}_\nu^{V}+ \mathbf{f}_\nu^B=0$,
where, for bead $\nu$, $\mathbf{f}^H_\nu$ is the hydrodynamic force,
$\mathbf{f}^{V}_\nu$ is the particle-particle excluded volume force,
$\mathbf{f}^{E}_\nu$ is the electrostatic force, $\mathbf{f}^B_\nu$ is the Brownian force
and $\mathbf{f}^S_\nu$ are any other potential forces that may arise in the system~\cite{hernandez-tps,Kounovsky-Shafer2013}.
The dynamics of the discrete ions in the solvent is described by the probability density
in configuration space.
The diffusion equation for the configurational distribution function has the form of a Fokker-Planck
equation, which corresponds to the following system of stochastic differential equations of motion
for the discrete ion positions:
\begin{equation}
\mathit{d}\mathbf{R} = \left[\mathbf{V}_0 + \frac{1}{k_BT}\mathbf{D}\cdot\mathbf{F}+
\frac{\partial}{\partial\mathbf{R}}\cdot\mathbf{D}\right]\mathit{dt} +
\sqrt{2}\mathbf{B}\cdot\mathit{d}\mathbf{W}.
\label{eq:bd}
\end{equation}
Here, ${\bf R}$ is a vector containing the $3N_P$ coordinates of the soft-particles,
and where ${\bf R}_\nu = {\bf x}_\nu$ denotes the Cartesian coordinates of particle
$\nu$. The vector $\mathbf{V}_0$, of length $3N_P$, represents the unperturbed velocity field,
with ${\bf V}_{0,\nu} = {\bf v}_0({\bf x}_\nu)$.
The vector ${\bf F}$ has length $3N_P$, with ${\bf F}_\nu = {\bf f}_\nu$ being the total non-Brownian,
non-hydrodynamic force acting on bead $\nu$. Finally, the independent
components of $d{\bf W}$ are obtained from a real-valued Gaussian
distribution function with zero mean and variance $dt$.
The diffusion tensor ${\bf D}$ (or mobility tensor) is a $3N_P\times 3N_P$
tensor. It may be separated into the Stokes drag
and the hydrodynamic interaction tensor, $\mathbf{\Omega}_{\nu\mu}$:
${\bf D}_{\nu\mu}= k_BT\left[ \boldsymbol{\delta}\delta_{\nu\mu} + (1-\delta_{\nu\mu})\mathbf{\Omega}_{\nu\mu} \right]$,
where $\boldsymbol{\delta}$ is a $3\times 3$ identity matrix and $\delta_{\nu\mu}$ is the Kronecker delta.
The Brownian perturbation, $d{\bf W}$, is coupled to the hydrodynamic interactions through
the fluctuation-dissipation theorem: ${\bf D} = {\bf B}\cdot{\bf B}^T$.

The characteristic variables for the system are set by the soft-particle:
hydrodynamic radius, $a$, for length, particle diffusion time, $\zeta a^2/k_BT$, for time (where
$\zeta=6\pi\eta a$ is the drag coefficient),
$e/4\pi\epsilon_0\epsilon a$ for the electrostatic potential, and the elementary charge $e$ for the charge.
There are two scales for velocity: one for the unperturbed velocity field $v_0$ and one for the
velocity fluctuations $u_c = k_BT/\zeta a$.
The uniform concentration for one of the species, $C_0$, is used as the characteristic concentration
of ions within the solvent. Therefore,
$P_{e,j}=v_0a/D_j$ defines a Peclet number for species $j$ based on the imposed flow field $\mathbf{v}_0$ and
$\psi_j= \zeta D_j/k_BT$ is the ratio between particle and continuum-ion diffusion coefficients.
The ratio between electrostatic forces and thermal forces defines the so-called Bjerrum length,
$\lambda_B = e^2/4\pi\epsilon_0\epsilon k_BT$, and the ionic
strength, $I=\tfrac{1}{2} \sum_{j}^{N_I}C_jz_j^2$, defines the so-called Debye length,
$\lambda_D^{-2} = 2N_Ae^2I/\epsilon_0\epsilon k_BT$.

% NP-GgEm
The key feature of GgEm-like methods is to decompose the charge-density and force-density expressions into a local (free-space) 
contribution and a global (bounded) contribution, in analogy to Ewald-like methods.
The ``local" densities are defined by
\begin{equation}
\begin{split}
\rho_l({\bf x}) & = \sum_{\nu=1}^{N_P} z_\nu\left[ \delta({\bf x-x}_\nu)-g_E({\bf x-x}_\nu)\right], \\
\boldsymbol{\rho}_l({\bf x}) &= \sum_{\nu=1}^{N_P} {\bf f}_\nu \left[\delta({\bf x-x}_\nu)-g_H({\bf x-x}_\nu)\right],
\end{split}
\end{equation}
which produces a local contribution to the electrostatic potential $\phi_l({\bf x})$ and
the velocity perturbation ${\bf u}_l({\bf x})$. The ``global" densities are given by
\begin{equation}
\begin{split}
\rho_g({\bf x}) & = \sum_{j=1}^{N_I} z_j \beta_j C_j + \sum_{\nu=1}^{N_P} z_\nu \left[g_E({\bf x-x}_\nu)\right], \\
\boldsymbol{\rho}_g({\bf x}) &= \lambda_b \sum_{j=1}^{N_I} z_j \beta_j C_j{\bf E} +
\sum_{\nu=1}^{N_P} {\bf f}_\nu\left[ g_H({\bf x-x}_\nu) \right],
\end{split}
\end{equation}
and are responsible for the global contribution of the potential $\phi_g({\bf x})$ and velocity
perturbation ${\bf u}_g({\bf x})$.
The linearity of the Poisson and Stokes equations implies that
$\phi({\bf x}) = \phi_l({\bf x}) + \phi_g({\bf x})$ and ${\bf u}({\bf x}) = {\bf u}_l({\bf x}) + {\bf u}_g({\bf x})$.
The screening functions, $g_{E,H}({\bf x})$, satisfy $\int_{\text{all space}} g_{E,H}({\bf x}) d{\bf x} = 1$.
The local contributions, $\phi_l({\bf x})$ and ${\bf u}_l({\bf x})$, are calculated analytically assuming an unbounded domain, only considering neighbor particles.
For Poisson's equation, the screening function is a Gaussian: $g_E(r)=(\alpha^3/\pi^{3/2})e^{(-\alpha^2 r^2)}$. For the Stokes equations it is a modified Gaussian~\cite{hernandez-ggem}:
$g_H(r) = (\alpha^3/\pi^{3/2})e^{(-\alpha^2r^2)}[5/2-\alpha^2r^2]$
(The local calculation scheme is described in the SI).
The global contributions are found numerically, requiring that
$\phi_{l}+\phi_g$ and ${\bf u}_l +{\bf u}_g$ satisfy appropriate boundary conditions.
For example, homogeneous Dirichlet boundary conditions require
$\phi_{g}({\bf x}) = -\phi_{l}({\bf x})$ and ${\bf u}_{g}({\bf x}) = -{\bf u}_{l}({\bf x})$.
For problems with periodic boundary conditions, Fourier techniques are
used to guarantee the periodicity of the global contributions. The
periodicity for local contributions is obtained through a minimum image convention.
A numerical implementation for periodic domains is described in the SI.

%Regularization
The point-source regularization (i.e. soft-particle) is implemented with the same screening functions from GgEm introducing two 
additional length scales $\xi_E^{-1}$ and $\xi_H^{-1}$. These length scales are related to the hydrodynamic radius $a$.
For the point-force (hydrodynamics), this is achieved by limiting the maximum
velocity on the fluid driven by the regularized-force, which at $Re=0$ is given by Stokes' law.
The regularization for the point-charge is achieved by distributing the total charge throughout the particle, thereby
``confining" the charge to the particle volume. The regularization parameters are
$\xi^R_E = 3/a$ and $\xi^R_H = \sqrt{\pi}/3a$.

\begin{figure}
\begin{center}
\includegraphics[width=0.48\textwidth]{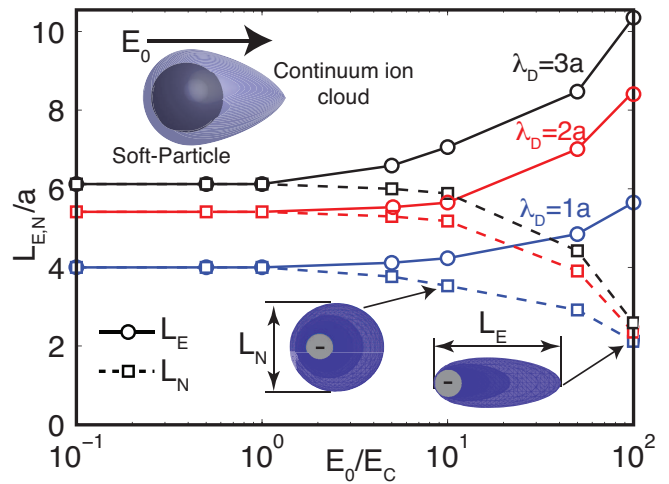}
\caption{Ionic cloud dimensions, in the neutral ($L_N$) and applied field direction ($L_E$), as a function of the
applied field for different Debye lengths (salt concentration). The soft-particles are embedded in a charged solvent and
the applied field deforms the continuum ionic cloud. Blue iso-concentration surfaces depict the counter-ions 
surrounding the negatively charged soft-particles. }
\label{fig:system}
\end{center}
\end{figure}

% Results
The method is validated by relying on several approximate analytical solutions for soft-particles.
Most of these solutions correspond to equilibrium conditions, and result
from the linearized Poisson-Boltzmann approximation (i.e. low potentials). Solutions for the
electrostatic potential and the electrostatic energy between particles are included in the SI.
For instance, Fig.~\ref{fig:system} illustrates how ionic clouds that surround the 
soft-particles are deformed when an external electric field is applied, as a function of the 
salt concentration. 

\begin{figure}
\begin{center}
\includegraphics[width=0.48\textwidth]{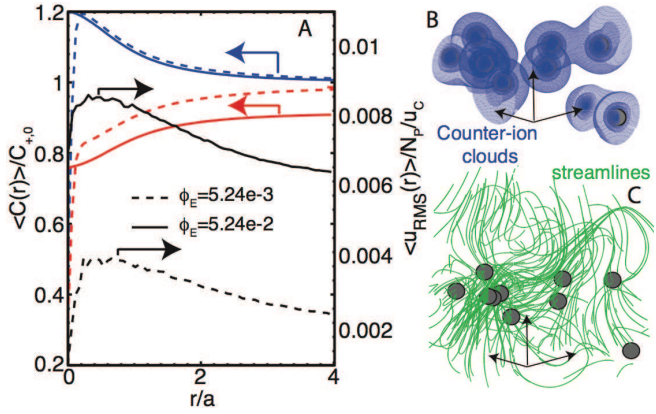}
\caption{(A) Average concentration and velocity fluctuations as a function of the distance from the soft-particle, $r/a$. 
Red and blue lines are for co- and counter-ion concentration profiles, respectively; black lines depict RMS velocity profiles. 
(B) Instantaneous counter-ion clouds (blue) and (C) velocity streamlines (green). 
The soft-particles have negative changed and they are embedded in a solvent with $\lambda_D=2a$ at an effective volume 
fraction $\phi_\text{e}=5.24\times10^{-3}$ ($N=10$). }
\label{fig:10}
\end{center}
\end{figure}

Non-equilibrium, finite concentrations, interacting ionic clouds and fluctuating hydrodynamic
interactions are built in the NP-GgEm. To illustrate the behavior of these types of systems, Brownian soft-particles are 
simulated in a solvent at different $\lambda_D$s (salt concentrations).
Diffusing soft-particles of charge $z_\nu=-1$ and $z_\nu=-2$ are simulated in a periodic box of size $L=20a$. 
Note that global electroneutrality is enforced through the continuum ions.
Soft-particle concentration is defined in terms of an effective volume fraction: $\phi_\text{E} = 4N_p\pi a^3/3L^3$. 
Figure~\ref{fig:10} provides representative results for NP-GgEm. In the figure,
average counter- and co-ion concentration and RMS velocity profiles are shown as a 
function of distance from the soft-particle, for $\lambda_D=2a$. At equilibrium, 
the 2D concentration profiles suggest that they follow PB-like behavior.
However, instantaneous concentration iso-surfaces (ion clouds) demonstrate  how, even at low concentrations, 
the ion clouds merge and interact with each other, being deformed by the Brownian motion of the particles.
The figure also includes the RMS velocity perturbation (normalized by the number of particles) as a function of the distance from the 
soft-particle at to different concentrations. The flow streamlines (green) are also plotted. 
The RMS velocity exhibits a slow decaying tail, accounting for the long-range behavior of HI, and
it increases non-linearly with concentration.

\begin{figure}
\begin{center}
\includegraphics[width=0.48\textwidth]{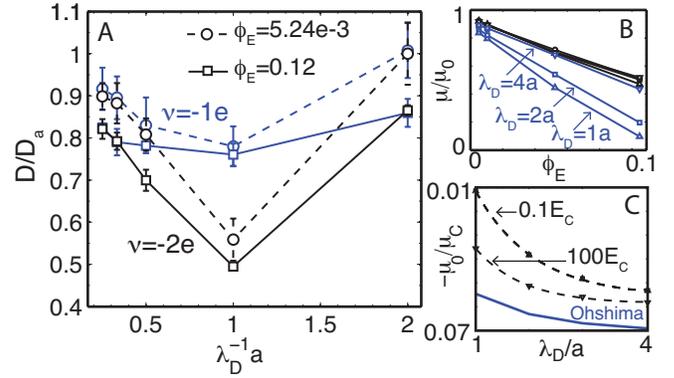}
\caption{(A) Diffusion coefficient as a function of the salt concentration (Debye length, $\lambda_D$), 
soft-particles charge, $\nu$, and effective volume fraction, $\phi_E$. 
The diffusion coefficient is normalized with the particle diffusion coefficient
$D_a =k_BT/6\pi\eta a$. 
(B) Electrophoretic mobility, $\mu$,  as a function of the effective volume fraction and salt concentration for
a weak (blue lines) and a strong (black lines) imposed fields, $E_0$. 
(C) Electrophoretic mobility at zeroth concentration, $\mu_0$, as a function of the Debye length for the two
applied fields in (B). The blue line represent Ohshima's approximate solution for non-fluid-penetrated 
soft-particles~\cite{Ohshima2012}. }
\label{fig:difusion}
\end{center}
\end{figure}

One key property of interest in charged colloidal systems is the diffusion coefficient,
and its dependence on the concentration, charge and salt. 
To illustrate the behavior of this property, and the usefulness of the approach proposed here, in 
Fig.~\ref{fig:difusion} we show the soft-particle diffusion coefficient as a function of the Debye length for 
two different concentrations and charges. 
The simulations were evolved from 200 to 1000 characteristic particle-diffusion times, thereby providing sufficient statistics. 
In general, the diffusion coefficient decreases by increasing soft-particle concentration and charge.
For high salt (low $\lambda_D$), the diffusion coefficient approaches the infinite dilute and non-charge system. 
As the salt concentration increases, electrolyte drag and collective motion drive a non-monotonic behavior, decreasing 
mobility at first and then increasing it. This behavior of diffusing charge particles in a charge solvent was also been
observed in other systems~\cite{Schumacher:1991ku,VizcarraRendon:1990ua,Banchio:2008un}. 
In the figure, the electrophoretic mobility, $\mu= u/E_0$, for soft-particles of charge $\nu=-1$ is also shown. The figure
includes two applied electric fields: a weak $E_0=0.1E_C$ and a strong $E_0=100E_C$ (where the characteristic
field is $E_C = e/4\pi\epsilon_0\epsilon a^2$). We recall that fluid can penetrate the soft-particles,
causing the electrophoretic mobility to change with the applied electric field. 
The electrophoretic mobility at zero concentration, $\mu_0$, and an approximate solution for non-fluid-penetrated soft-particles~\cite{Ohshima2012}
as a function of the Debye length for the two applied fields are both included in the figure.  For weak fields,
there is a stronger dependence on Debye length, and the electrophoretic mobility approaches to zero, meaning that the 
fluctuating hydrodynamic interactions (driven by Brownian motion) affect particle trajectories, averaging down the 
electrophoretic induced motion. On the other hand, for strong fields, the normalized mobility collapses into a single curve,
maintaining the dominance of the electrokinetic forces. Ohshima's solution is obtained from the linearized 
PB; other authors had suggested different solutions that result in similar approximations for  
these electrokinetic properties, which are restricted to zero concentrations and weak potentials.
The approach we proposed here avoids the approximation of electrokinetic properties and relationships. 
It provides a platform for other colloidal and polymeric systems, including deformable objects, rigid suspensions and cells, where
lubrication forces and slip conditions are important. 

With the method described here, it is possible to perform efficient simulations of discrete charged elements 
embedded in charged solvents at non-equilibrium conditions.  
It should find applications in a wide variety of situations involving the physics of polymeric and colloidal systems 
including micro- and nano-scale systems. 

\bibliographystyle{apsrev4-1}
\bibliography{referencias}

\end{document}